\documentclass[pdftex,twocolumn,epjc3]{svjour3}          

\usepackage[T1]{fontenc}
\usepackage[english]{babel}
 \addto\captionsenglish{
                        \renewcommand{\figurename} {FIG.}%
                        \renewcommand{\tablename}  {TABLE}%
                        \renewcommand{\appendixname}{APPENDIX}%
                       }
\usepackage[numbers,sort&compress]{natbib} 

\usepackage{graphicx}
 \graphicspath{{graphics/}}
\usepackage{multirow}

\usepackage[caption=false]{subfig}  
\usepackage[export]{adjustbox}

\usepackage{amsmath}
\usepackage{amssymb}
\usepackage[version=3]{mhchem}
\usepackage{mathrsfs}
\usepackage{bm}
\usepackage{todonotes}
\usepackage{upgreek}
\usepackage{soul} 
\usepackage{doi} 
\hypersetup{
            colorlinks,
            linkcolor={red!70!black}  ,
            citecolor={green!70!black},
            urlcolor ={blue!70!black}
           }

\usepackage{comment} 
\usepackage[switch]{lineno} 
\usepackage{xspace}

\begin{document}

 \journalname{Eur. Phys. J. C}
 \title{
 First measurement of GaAs as a scintillating calorimeter: achievements and prospects} 

 \author{
            A.~Melchiorre~\thanksref{INFNlngs,UnivAQ,corrAuth} 
         \and D.~L.~Helis~\thanksref{INFNlngs,corrAuth} 
         \and  A.~Puiu~\thanksref{INFNlngs,corrAuth}
         \and G.~Benato~\thanksref{GSSI, INFNlngs}
         \and P.~Carniti~\thanksref{Mib}
         \and I.~Colantoni~\thanksref{CNRroma,INFNroma}
         \and A.~Continenza~\thanksref{UnivAQ}
         \and N.~Di Marco\thanksref{GSSI,INFNlngs}
         \and A.~Ferella~\thanksref{UnivAQ,INFNlngs}
         \and C.~Ferrari~\thanksref{GSSI,INFNlngs}  
         \and F.~Giannessi~\thanksref{UnivAQ}
         \and C.~Gotti~\thanksref{Mib}
         \and E.~Monticone~\thanksref{INRIM}
         \and S.~Nagorny~\thanksref{GSSI,INFNlngs}
         \and E.~Olivieri~\thanksref{CNRS}
         \and L.~Pagnanini~\thanksref{GSSI,INFNlngs}     
         \and G.~Pessina~\thanksref{Mib}
         \and C.~Petrucci~\thanksref{GSSI,INFNlngs}
         \and S.~Pirro~\thanksref{INFNlngs}
         \and A.~Prajapati~\thanksref{INFNlngs,UnivAQ}
         \and G.~Profeta~\thanksref{UnivAQ}
         \and M.~Rajteri~\thanksref{INRIM}
         \and P.~Settembri~\thanksref{UnivAQ}
         \and A.~Shaikina~\thanksref{GSSI,INFNlngs}
         \and C.~Tresca~\thanksref{CNR}                
         \and D.~Trotta~\thanksref{Mib}   }
    \institute{INFN -- Laboratori Nazionali del Gran Sasso, Assergi, I-67100 L'Aquila, Italy\label{INFNlngs}
            \and Dipartimento di Scienze Fisiche e Chimiche, Universit\`a degli Studi dell'Aquila, I-67100 L'Aquila, Italy\label{UnivAQ}
            \and Gran Sasso Science Institute, I-67100 L'Aquila, Italy\label{GSSI}
            \and INFN and Università degli studi di Milano-Bicocca, I-20126 Milano, Italy \label{Mib} 
            \and Consiglio Nazionale delle Ricerche, Istituto di Nanotecnologia, c/o Dip. Fisica, Sapienza Università di Roma, 00185, Rome, Italy\label{CNRroma}
            \and INFN, Sezione di Roma, P.le Aldo Moro 2, 00185, Rome, Italy\label{INFNroma}
            \and Université Paris-Saclay, CNRS/IN2P3, IJCLab, 91405 Orsay, France\label{CNRS}
            \and Istituto Nazionale di Ricerca Metrologica Torino, I-10135 Torino, Italy\label{INRIM} 
            \and CNR-SPIN, Università degli studi dell’Aquila, I-67100 L’Aquila, Italy\label{CNR}
            \\ 
         }
           \newcommand{\emailaddress}{andrei.puiu@lngs.infn.it, \\ andrea.melchiorre@lngs.infn.it, dounia.helis@lngs.infn.it}
           \thankstext{corrAuth}{Corresponding Authors: \href{mailto:\emailaddress}{\emailaddress}}

 \twocolumn
 \maketitle
 \begin{abstract}

In this paper we present the first measurement of a Gallium Arsenide (GaAs) crystal as a scintillating calorimeter with dual heat and light readout within the DAREDEVIL project. 

The experimental setup features a 4.3 g GaAs (GaAs-1) crystal, operated at approximately 10 mK coupled with a Neutron Transmutation Doped (NTD) thermal sensor for phonon detection and an auxiliary calorimeter for the detection of scintillation light. For the GaAs-1 crystal, a baseline resolution of 121 ± 2 eV has been achieved. While, with a 3.5 g GaAs (GaAs-2) crystal an even better baseline resolution of 44.5 $\pm$ 0.8 eV was achieved.

Alpha and X-ray calibration sources were used to study the scintillation light response to different types of interacting radiation. The GaAs crystal exhibits a strong particle discrimination capability based on the emitted scintillation light, featuring a light yield (LY) of 0.9 $\pm$ 0.2  keV/MeV for $\alpha$ induced events and 0.07 $\pm$ 0.01 keV/MeV for $\beta$/$\gamma$ events, both measured at 1~MeV.

The unusual luminescence behavior, i.e. more light being produced under irradiation by $\alpha$ particles warrants further investigation, particularly due to its potential to enhance sensitivity to low-energy nuclear recoils from light dark matter scattering.

 \end{abstract}
 

 \section{Introduction}

The search for Dark Matter (DM), which constitutes approximately 27$\%$ of the universe’s mass, has driven innovative detection strategies in particle physics. Recently, one promising experimental approach was introduced for the direct DM detection through DM-electron scattering mechanism, which enables the exploration of sub-GeV DM candidates beyond traditional nucleon-based interactions~\cite{PhysRevD.100.075028}.

Indeed, current efforts on a direct DM detection are primarily focused on Weakly Interacting Massive Particles (WIMPs) in the GeV–TeV range, with multi-ton-scale experiments such as LZ~\cite{LZ:2023poo}, XENONnT~\cite{PhysRevLett.131.041003}, DarkSide~\cite{Rossi:2024lae}, and PANDA-X~\cite{PandaX-II:2021nsg} pushing towards the so-called "neutrino floor"~\cite{PhysRevLett.127.251802}. However, sub-GeV dark matter has gained increasing attention, with models such as asymmetric dark matter~\cite{PhysRevD.79.115016}, freeze-in~\cite{Hall:2009bx}, and strong dynamics~\cite{PhysRevLett.113.171301} expanding the search window down to eV-scale DM candidate particles~\cite{XENON:2020rca}.
The interaction between dark matter and electrons is particularly significant as it allows to access to lower mass DM candidates compared to the coherent nucleus scattering~\cite{Billard:2021uyg}. Recent results from XENONnT~\cite{XENON:2022ltv,xenon2024} have placed stringent constraints on DM-electron scattering cross-sections, demonstrating the power of electronic recoil detection in probing new parameter regions. Other experiments, including LUX~\cite{PhysRevLett.122.131301}, PANDA-X~\cite{PandaX-II:2021nsg}, and DarkSide-50~\cite{Franco:2023sjx}, have also shown sensitivity to sub-GeV dark matter, while CRESST~\cite{Cresst}, DAMIC-M~\cite{DAMIC-M:2023gxo}, EDELWEISS~\cite{EDELWEISS:2020fxc}, SENSEI~\cite{SENSEI:2023zdf}, and SuperCDMS~\cite{SuperCDMS:2018mne} have extended limits into the MeV/$c^2$ mass region.
In this context, the DAREDEVIL project
aims to develop a low-energy detection threshold, high-resolution cryogenic detector for future light DM candidates searches. The approach involves coupling state-of-the-art cryogenic sensors to materials with small or zero band gaps, such as semiconductors and Dirac/Weyl crystals~\cite{Derenzo:2016fse,Derenzo:2018plr,Vasiukov:2019cbf}, and operating them as cryogenic calorimeters~\cite{Pirro:2017ecr}. 
Theoretical models suggest that low-mass DM may scatter off electrons, producing detectable electron recoils~\cite{Billard:2021uyg}. Extensive theoretical calculations~\cite{PhysRevD.85.076007} supports the potential of semiconductor-based detectors in uncovering interactions involving DM particles below the MeV/$c^2$ mass scale.
A promising detector material for DM candidates in this mass region should combine several crucial properties:
\begin{enumerate}
    \item use of a polar crystal, that would enhance sensitivity to dark photon absorption~\cite{Knapen:2017ekk}.
    \item use of a semiconductor material, allowing for controlled and effective charge collection.
    \item use of a scintillating material with an $\mathcal{O}(\text{eV})$ band gap (significantly smaller than NaI (5.86~eV) or CsI (3.5 eV))~\cite{PhysRevD.96.016026} that will significantly increase the number of light photons emitted and improve the particle identification capability.   
\end{enumerate}

Gallium Arsenide (GaAs) is a particularly promising detector material due to its unique combination of physical and electronic properties, especially when used as a calorimeter at milliKelvin (mK) temperatures. Its direct band gap of approximately 1.42~eV at room temperature is crucial for achieving high sensitivity to low-energy excitations, enabling the detection of sub-MeV DM particles through electron recoil processes. Moreover, GaAs is predicted to have a higher DM-electron fiducial cross-section compared to other low band gap semiconductors and metals~\cite{PhysRevD.96.016026}. It also outperforms other detector materials based on insulators for specific types of DM interactions, such as scattering mediated by dark photons or dark photon absorption~\cite{Zema:2024epe, Knapen:2021bwg}.
As a polar material, GaAs is predicted to be particularly sensitive to DM interactions that involve optical phonons, dark photon mediators, or the Migdal effect~\cite{PhysRevD.105.015014, Knapen:2021bwg, Essig:2022dfa}. 
 
The pure (undoped) GaAs crystal exhibits also a decent scintillating properties, producing 2~photons/keV under X-ray irradiation at 10~K~\cite{Vasiukov:2019cbf}. Furthermore, recent studies demonstrate that GaAs scintillation properties could be further adjusted and enhanced through proper dopant selection~\cite{Vasiukov:2019cbf, Griffin:2018bjn}. Thereby, the GaAs material allows the implementation of a dual readout system, where both scintillation light and phonon signals are recorded simultaneously. 

Indeed, to study the scintillation light one can use a Germanium light detector (Ge-LD), which consists of a thin (about 0.5~mm) high-purity germanium monocrystalline slabs, see for instance~\cite{pirro2006development,beeman2013characterization}. Scintillation photons interact with the LD and deposit their energy within its thin surface layer. Absorbed energy is subsequently converted into phonons, leading to a temperature increase of the system.
Moreover, in order to further increase the sensitivity to the scintillation light, one can exploit the Neganov-Trofimov-Luke effect (NTL) on the Ge-LD. The NTL effect consists of the acceleration of primary electrons/holes produced under scintillation light and their further multiplication under the external bias (at the range of 50--100~V) that allows to amplify the phonon signal in Ge-LD without increasing the noise, i.e. improving the signal-to-noise ratio and lowering the energy threshold.
Hence, applying a bias voltage to the Ge-LD enhances the particle discrimination capability and improves the sensitivity to rare events and weak light signals~\cite{Neganov:1985khw,Luke:1988xcw}. 
Therefore, this dual-readout system, combining phonon detection in GaAs and scintillation light measurement in Ge-LD, can provide a robust event-by-event particle identification and minimises the risk of misinterpreting background events. 

In the following paragraphs, we present the first measurements of scintillation properties of GaAs at cryogenic temperature (around 10~mK) illuminated by X-rays and $\alpha$ particles in a wide energy range, the first test of the GaAs crystal as a dual-readout scintillating calorimeter, and an improved performance of GaAs in a phonon channel that allows to significantly enhance sensitivity to low-mass DM candidates.

 \section{Experimental Setup} 
\begin{figure}
    \centering
    \includegraphics[width=0.45\textwidth]{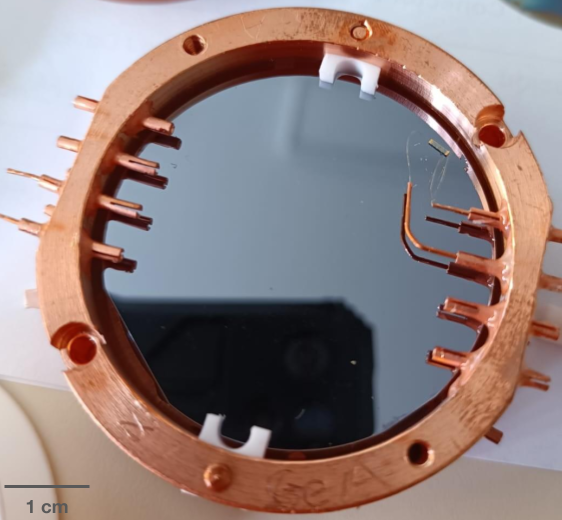}
    \caption{The GaAs-1 detector mounted with PTFE clamps on the copper frame. On the right-hand side, the Ge-NTD thermistor is glued and connected via two 25-$\mu$m-thick gold wires to readout the heat signal. The whole frame was then coupled to the Ge-LD frame and to the 10~mK stage of the dilution refrigerator.}
    \label{fig:set}
\end{figure}

Two GaAs single crystal wafers purchased from Sigma were used as scintillating calorimeters in the present studies. Both wafers are 2-inch in diameter, but differ slightly in thickness - 0.5~mm (4.3~g, GaAs-1) and 0.325~mm (3.5~g, GaAs-2). The wafers were equipped with a 3$\times$0.6$\times$0.4~mm Ge-NTD~\cite{ntd} sensor using three epoxy glue spots onto the polished surface of the GaAs crystals. The GaAs wafers were then mounted with polytetrafluoroethylene (PTFE) clamps on a copper frame, as shown in Fig.~\ref{fig:set}.

 \begin{figure}
    \centering
    \includegraphics[width=0.45\textwidth]{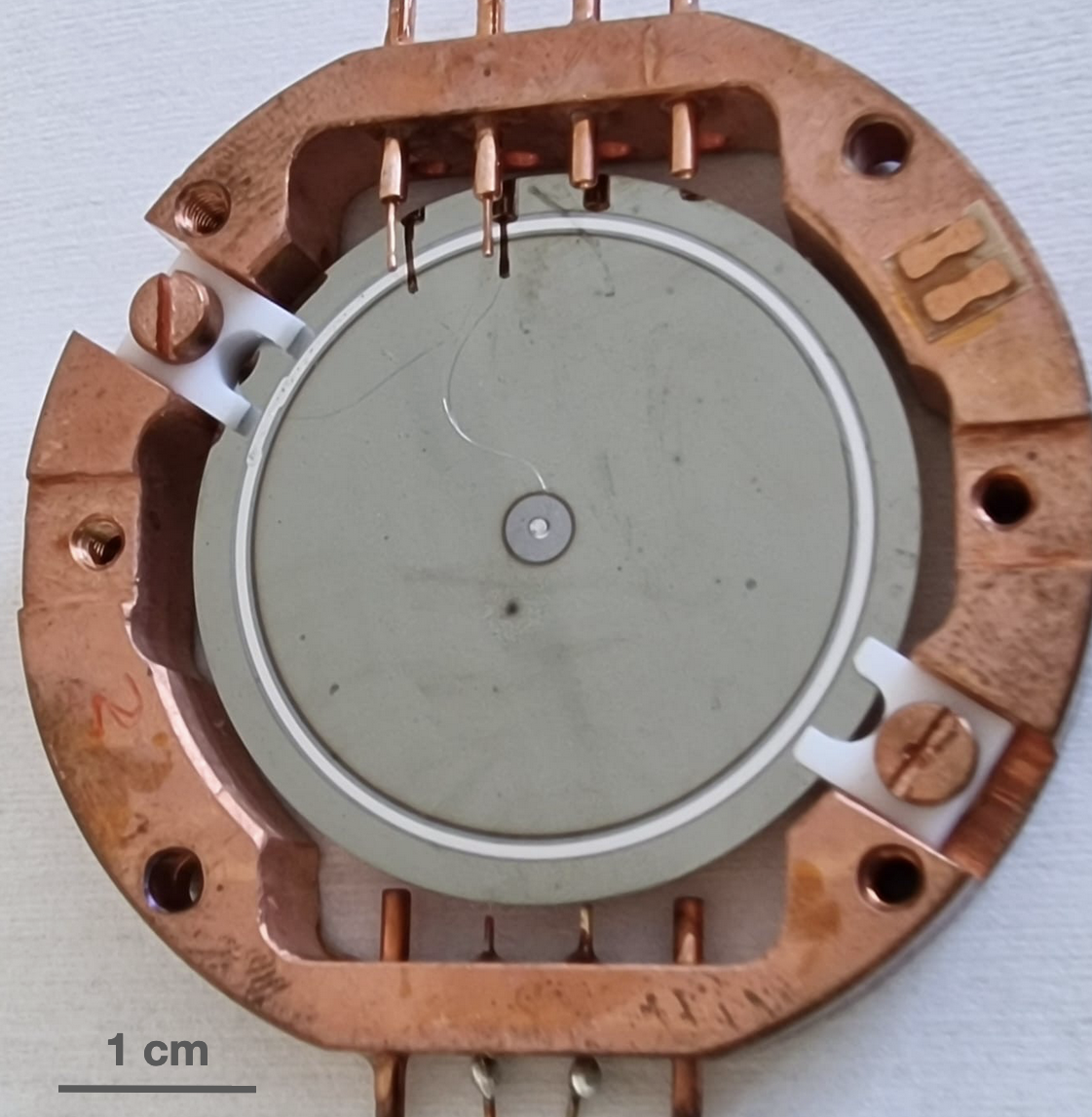}
    \caption{The Ge light detector (Ge-LD) mounted with PTFE clamps on the copper frame. The image shows the electrodes for the NTL-effect, while the Ge-NTD thermistor is mounted on the opposite face and bonded with 25~$\mu$m gold wires, which were connected to copper pins. The entire frame was then coupled to the GaAs detector frame and to the 10~mK stage of the dilution refrigerator.}
    \label{fig:set2}
\end{figure}

Additionally, we used a Germanium wafer as a light detector (Ge-LD) with the same type of Ge-NTD sensor as the GaAs wafers, attached to the surface of a Ge crystal in the same manner. The Ge-LD was mounted on a separate copper frame, fixed on the top of the GaAs detector frame, with a 10 mm separation between the two wafers. The Ge-LD electrodes, configured in a dot-circle pattern, for the NTL effect were bonded with 25~$\mu$m gold wires, which were connected to copper pins as shown in Fig. ~\ref{fig:set2}. 
 
 Both detectors (GaAs and Ge-LD) were then thermally linked to the mixing chamber of a pulse-tube assisted dilution refrigerator (IETI)~\cite{ieti}, located underground in Hall~C at the Laboratori Nazionali del Gran Sasso (LNGS, Italy), and operated at a base temperature of 10~mK. Although the detector setup was operated underground at LNGS, it is important to emphasise that no specific selection of low-radioactivity materials was performed.
 
The electrical connection for the detector and for the bias voltage to the cryostat is established by connecting copper pins with constantan twisted wires to the thermalised board on the mixing chamber. The signal from the Ge-NTD's is then read out by an electronics unit positioned at room temperature.

The room temperature readout electronics consist of low-noise DC-coupled front-end boards 
and high-resolution digitisers~\cite{Arnaboldi_2018,Carniti:2022coa}.
Signals from each detector were digitised with 24~bit ADC, followed by an anti-aliasing Bessel filter with a cutoff frequency of 500~Hz to eliminate higher-frequency noise, and synchronously recorded in a stream file using a MATLAB program~\cite{Carniti:2022coa}. \

In order to calibrate the detector energy response in the heat channel, a $\gamma$-ray source was employed, which consist of a Thorium tungstate (Th–W) wire, along with a $^{55}$Fe X-rays source to illuminate the GaAs wafer. Both calibration sources were positioned approximately 5~mm from the polished crystal surface, and have a total activity of approximately 1~Bq. This combination of $\gamma$-ray sources allows us to cover a wide range of energies.
Moreover, the detector setup was also equipped with a $^{238}$U/$^{234}$U calibration $\alpha$ source with a smeared energy profile, similar to used in~\cite{armatol2023zno,armatol2021cupid}, that allows to irradiate the GaAs wafer with a 4.2~MeV and 4.8~MeV $\alpha$ particles to characterise its light response with respect to $\alpha$ irradiation.

Similarly, for the energy calibration of the Ge light detector (Ge-LD), was used a $^{55}$Fe X-rays source with an activity of approximately 1~Bq, positioned at a 5~mm distance from the Ge-LD surface.

 \section{Data Analysis}
The first level data analysis have been processed via a custom-made suite, written in Octave, which
consisting of: \

1) triggering software, based on a simple band-pass
threshold algorithm, searching for the rising edge of the pulses recorded in the streaming. It saved the time postions of each triggred event.

2) Filtering processing program, based on the Gatti-Manfredi algorithm (Optimum filter OF) \cite{Gatti1986cw}. Each pulse window (defined around the trigger file positions) is
filtered in the frequency domain via the OF transfer function and reconstructed in the time-domain: the maximum of the time-domain filtered signal event is taken as the energy estimator of the impinging particle. The transfer function is built by combining a template pulse (obtained by averaging several pulses in order to obtain a final pulse with S/N about 100) 

with the average Noise Power Spectrum, this latter built by averaging windows where no pulse was triggered. 
For each event several parameters are evaluated, i.e. rise-time, decay-time, time domain pulse-filtered $\chi^2$, enabling the possibility to identify events with different shapes from the template pulse and identify the type of particle impinging the detector. Multi-channel synchronised  processing is made possible by using a common trigger file.

\section{Results}
\subsection{GaAs-1 crystal}
\begin{figure}[h]
    \centering
    \includegraphics[width=0.5\textwidth]{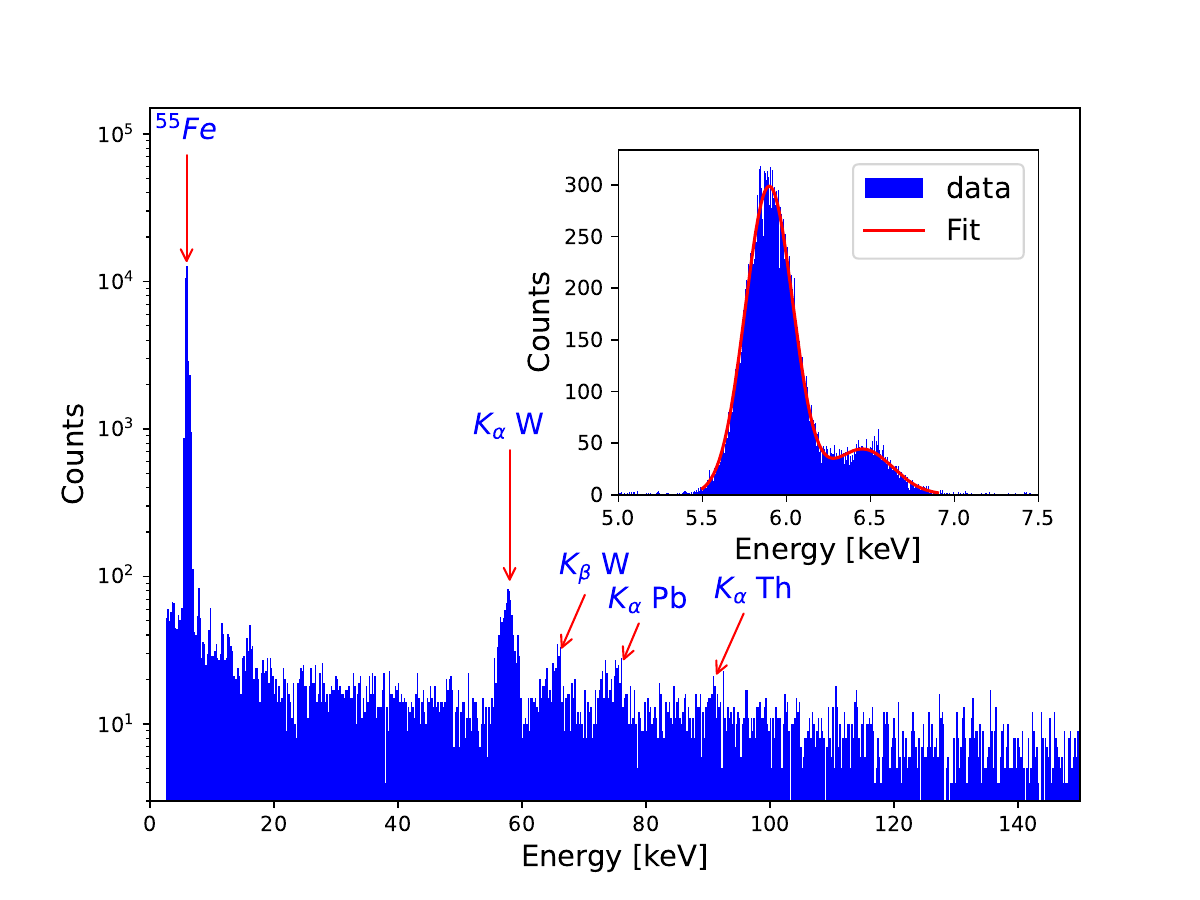}
    \caption{X-ray energy spectrum acquired with the 4.3~g GaAs detector, showing the $^{55}$Fe peak, the characteristic $K_{\alpha}$/$K_{\beta}$ lines of W, $K_{\alpha}$ lines of Pb and $K_{\alpha}$ lines of Th. The inset displays a zoomed view of the $^{55}$Fe prominently featuring the 5.9~keV Mn $K_{\alpha}$ line and 6.49~keV Mn $K_{\beta}$. }
    \label{fig:spectrumfe}
\end{figure}

With the 4.3~g GaAs detector a 17-hour-long calibration run was performed. Compared to our previous results~\cite{refId0}, we have improved the resolution of the GaAs detector in the heat channel by more than a factor of 2. Specifically, we can now clearly resolve the $K_{\alpha}$ and $K_{\beta}$ peaks of the $^{55}$Fe source as shown in inset of Fig. \ref{fig:spectrumfe}, achieving the energy resolution of $\sigma$~=~140~$\pm$~8~eV for the peaks. This value was extracted from the fit of the spectrum using the sum of two Gaussian functions, one for the Mn $K_{\alpha}$ peak and one for the $K_{\beta}$ peak, imposing the same sigma for both peaks. For the baseline resolution we obtained $\sigma$$_{RMS}$~=~121~$\pm$~2~eV.
This was made possible by improvements to the cryostat, including the addition of a spring for further mechanical decoupling and an enhanced pulse tube support to reduce vibrational noise. Furthermore, from Fig. \ref{fig:spectrumfe}, we can observe the $K_{\alpha}$ and $K_{\beta}$ peaks of W at 59.3 keV and 67.2 keV, respectively, as well as the $K_{\alpha}$ line of Th at 93.3 keV, all originating from the thoriated tungsten wire ${\gamma}$-ray source. In addition to these three lines, the $K_{\alpha}$ line of lead at 74.7 keV is also visible.

\begin{table}[ht!]
   \centering
    
    \begin{tabular}{l c c}

\hline 
    Mass & 4.3 & g \\ \hline
    Density & 5.32 & g/cm$^3$ \\ \hline
    Diameter & 5.08 & cm \\ \hline
    NTD response & 490 & $\mu$V/MeV \\ \hline
    Baseline resolution (RMS)  & 121 $\pm$ 2 & eV \\ \hline
    Peak $\sigma$ at 5.9 keV  & 140 $\pm$ 8 & eV \\ \hline
   \end{tabular}
   \caption{Summary of the performance of GaAs-1 detector operated as a low-temperature calorimeter.}
   \label{tab:perf}
\end{table}

\begin{figure}[h] 
    \centering
    \includegraphics[width=0.5\textwidth]{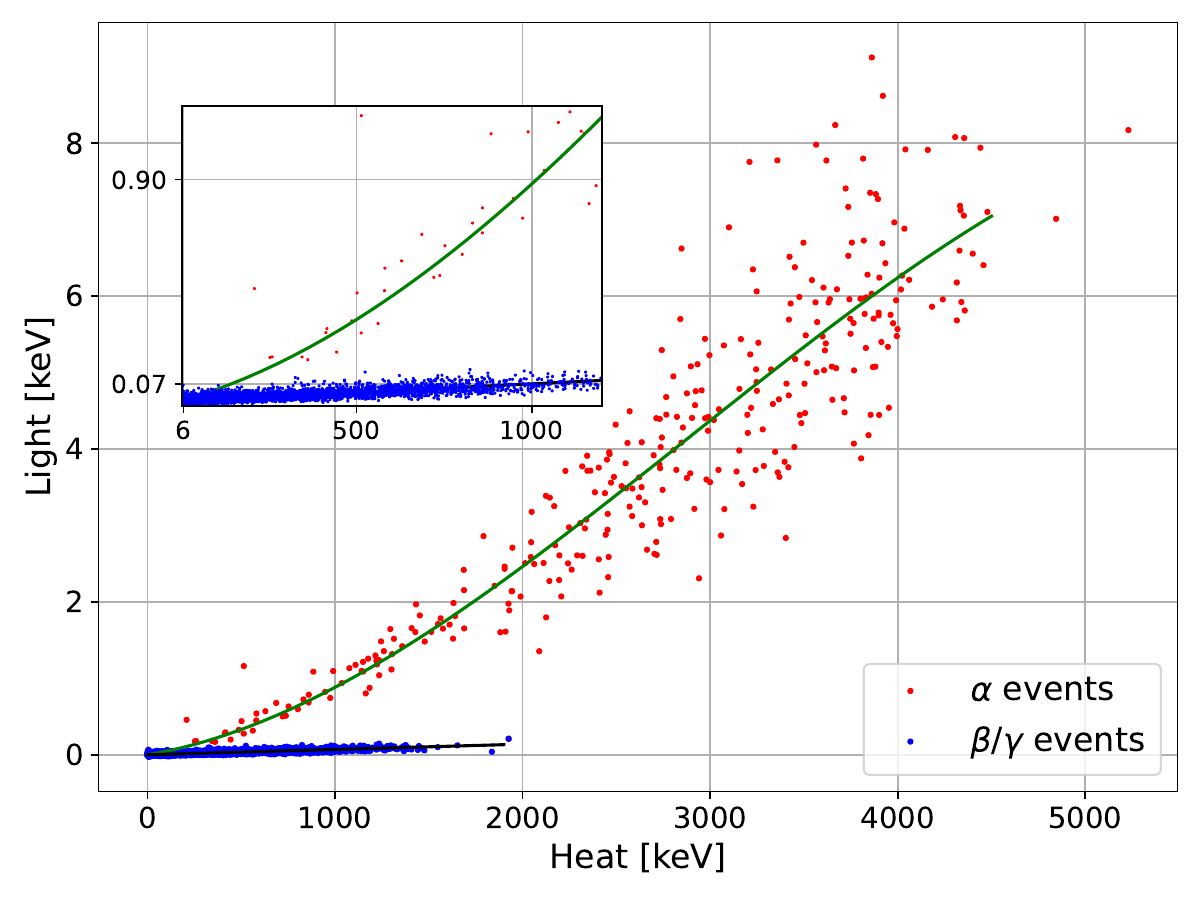}
    \caption{Light versus heat scatter plot constructed for events collected within the calibration run with the simultaneously installed $^{238}$U $\alpha$ source and X-rays from $^{55}$Fe and W. The population of $\beta$/$\gamma$ and $\alpha$ events is highlighted in blue and red, respectively. See more information in the text below. }
    \label{fig:scatterplot}
\end{figure}

Similarly, the Ge-LD data were processed to calibrate its energy spectrum using the $^{55}$Fe source. The Ge-LD without applying the NTL-effect exhibits a baseline resolution of $\sigma$$_{RMS}$~=~60.4~$\pm$~0.5~eV and a peak resolution of $\sigma$~=~102~$\pm$~2~eV. By exploiting the NTL-effect, the baseline resolution was improved by a factor of 12 reaching $\sigma$$_{RMS}$~=~5.2~$\pm$~0.1~eV.
Once the calibration of both detectors was completed, the data acquired with the NTL-effect enabled to study the scintillation light of GaAs. For each fired trigger in the heat channel, we simultaneously acquired the waveform in the Ge-LD data stream to investigate its corresponding signal.
This allowed us to record coincident signals, thus producing a light versus heat scatter plot including all detected events, as shown in Fig.~\ref{fig:scatterplot}.

\begin{table}[h]
    \centering
    \begin{tabular}{c|c}
        \hline
         & Light yield (keV/MeV) @ 1 MeV \\  
        \hline
        $\beta$/$\gamma$ & 0.07~$\pm$~0.01\\  
        $\alpha$ & 0.9~$\pm$~0.2 \\  
        \hline
    \end{tabular}
    \caption{Light yield values for $\beta$/$\gamma$ and $\alpha$ induced events.}
    \label{tab:light_yield}
\end{table}

This separation is made possible by the different scintillation light yields associated with each type of interaction.
From Fig. \ref{fig:scatterplot}, we observe a clear separation of the two different families. 
In the $\alpha$ band, between 2 and 4.5~MeV, events from the $^{238}$U/$^{234}$U source are observed. The $\alpha$ induced events appear smeared, and this can be explained by the use of a multilayer deposited source and by irradiation of the non-polished surface of the GaAs-1 wafer by $\alpha$ particles.

Scintillation events induced by $\alpha$ particles were fitted using a third-degree polynomial function ($E{_L}=a E{_H}^3+bE{_H}^2+cE{_H}$), in order to highlight the non-linearity in the light yield (LY) response to this type of interaction. In contrast, events belonging to the $\beta/\gamma$ band were fitted with a straight line ($E{_L}=a'E{_H}$). The fit coefficients obtained were $a = -0.06$(2), $b = 0.54$(1), and $c = 0.4$(2) for the $\alpha$ events, and $a' = 0.069$(1) for the $\beta/\gamma$ events. The light energy ($E{_L}$) is expressed in keV, while the heat energy ($E{_H}$) is expressed in MeV.

The zoom of the light versus heat scatter plot for the energy interval from 6~keV to 1.2~MeV energy deposition in the heat channel is shown as inset in Fig.~\ref{fig:scatterplot}. 
From this figure and using energy dependence of light for $\beta/\gamma$ events one can evaluate a number of photons emitted per keV, when GaAs excited by low-energy X-rays. Taking into account that emission maximum for the undoped GaAs crystal is 840~nm (1.476~eV)~\cite{Vasiukov:2019cbf} the derived LY is 0.05~photon/keV. 

Due to the absence of a clear and distinctive calibration $\alpha$ peaks of $^{238}$U/$^{234}$U and strong non-linear behavior of the light yield (LY) induced by $\alpha$ particles in a wide energy interval, to estimate the LY of the GaAs crystal, we decided to consider the amount of light corresponding to 1~MeV energy deposition in the heat channel. Moreover, it should be emphasised that because of the GaAs crystal geometry (a thin wafer weighing several grams) the energy scale for $\beta/\gamma$ events was calibrated using X-rays peaks up to 60~keV (from $^{55}$Fe and W) clearly visible in the energy spectrum, and assuming a linear detector response in the heat channel over a broader energy range.
The zoom of the light versus heat scatter plot up to 1.2~MeV energy deposition in the heat channel is shown as inset in Fig.~\ref{fig:scatterplot}, whilst the LY values for $\beta/\gamma$ and $\alpha$ events estimated at 1~MeV energy deposition are reported in the Tab.~\ref{tab:light_yield}. Additionally, it should be stressed that since the energy scale of the GaAs detector was calibrated with $\gamma$ sources and we cannot perform a proper energy calibration for $\alpha$ induced events because of no evident peaks, the real energy of $\alpha$ events represented at 1~MeV on Fig.~\ref{fig:scatterplot} can vary about 10--20\% from this nominal value. However, 
the possible mismatch of the nominal deposited energy for these $\alpha$ events is within the uncertainty of the LY value determination.

As one can see from Fig.~\ref{fig:scatterplot} and values listed in Tab.~\ref{tab:light_yield}, events induced by $\alpha$ particles characterised by a higher value of LY with respect to $\beta/\gamma$ events, by a factor of 10. This is not a typical LY behavior for scintillating crystals. With a few exceptions, typically the amount of light produced in interactions with $\alpha$ particles is lower than that for $\beta/\gamma$s of the same energy, and varies in the range from 0.2 to 0.6 of that for the $\beta/\gamma$ induced events, see~\cite{poda2021scintillation}. This reduction in scintillation light emission for $\alpha$ particles is described as a Quenching Factor (QF), and it is commonly explained on the basis of a saturation effect of luminescence centers due to the high ionisation density that characterises the interaction of a heavy charged particle in matter~\cite{birks1951scintillations,TRETYAK201040}. However, as mentioned above, there are several exceptions amongst scintillating materials, for which QF~>~1. The first amongst them is the ZnSe scintillating crystal for which QF~>~1, yet the QF value is hardly reproducible and varies widely from one sample to another, reaching QF~>~3--6 for some cases~\cite{arnaboldi2011characterization,beeman2013performances,artusa2016first,nagorny2017quenching}.

The second example is the ZnO crystal, for which in cryogenic measurements as a scintillating calorimeter, the LY for events induced by $\alpha$ particles spans between 0.2 to 3.0~keV/MeV and exceeds the amount of light produced by $\beta/\gamma$ events with a LY = 1.5(3)~keV/MeV~\cite{armatol2023zno}.

Similarly, the shape, intensity and energy position of the emission peaks, especially in the 800--900~nm wavelength interval, in GaAs crystals are reported to be strongly depended on the material chemical purity, type of dopants, defect structure, and on the type of conductivity, as shown in~\cite{Vasiukov:2019cbf} and references therein.  

In all above mentioned cases, such anomalous behavior of scintillation mechanism cannot be easily accommodated within the existing theoretical framework used to describe the scintillation properties of materials. Hence further investigations of this phenomenon, i.e. the light yield enhancement of $\alpha$ particles in GaAs crystal, are strongly required. Nevertheless, this feature is very important for particle discrimination, since when the light yield is a strong function of the particle type, the effective particle discrimination can be achieved.

\subsection{GaAs-2 crystal}
With the 3.5~g GaAs detector a 72-hour-long calibration run was performed, exposing the GaAs crystal only to a $^{55}$Fe X-ray source. Data acquisition and analysis were performed using the same software framework previously employed.

\begin{figure}[h]
    \centering
    \includegraphics[width=0.5\textwidth]{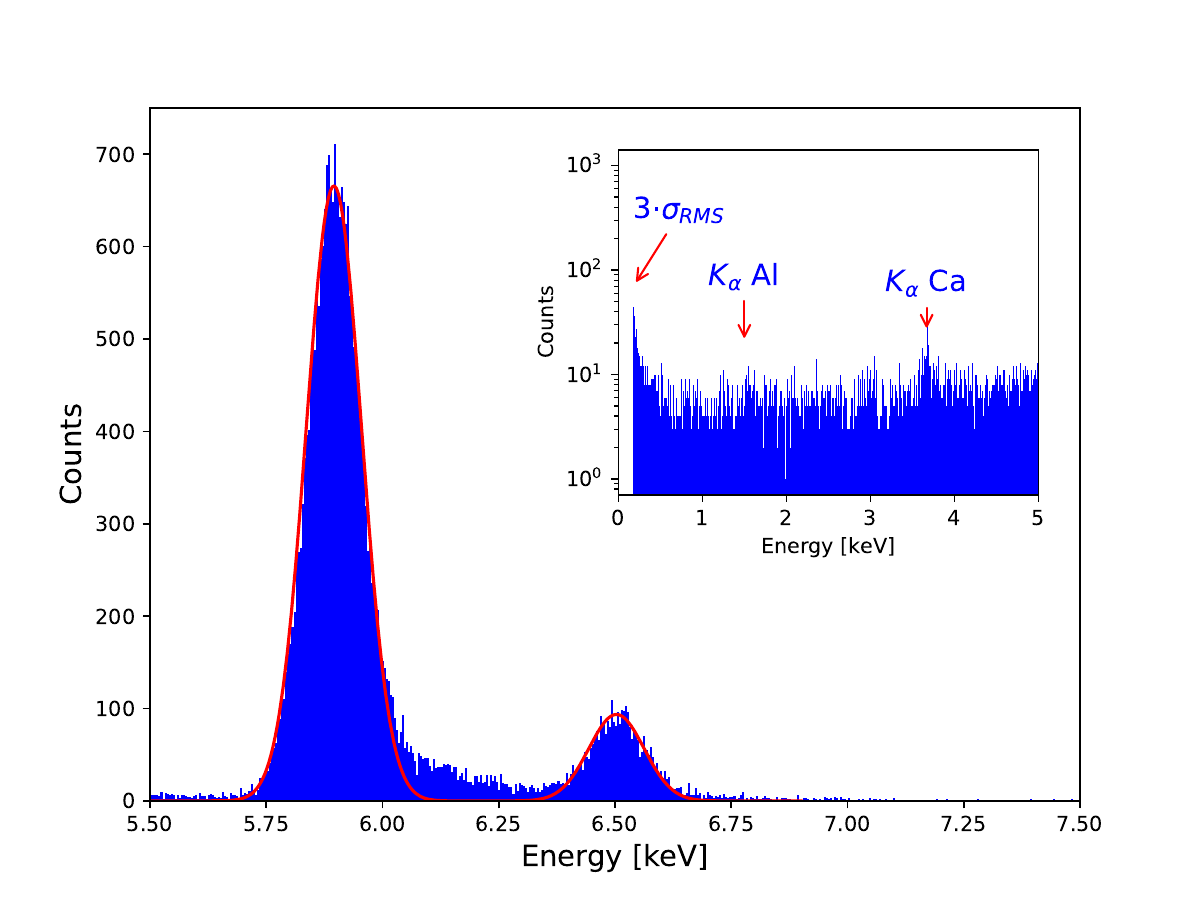}
    \caption{The low-energy X-ray spectrum acquired with a 3.5~g GaAs detector and prominently featuring the 5.9~keV Mn $K_{\alpha}$ line and 6.49~keV $K_{\beta}$. Notably, the energy resolution at the 5.9~keV Mn $K_{\alpha}$ line peak is measured at $\sigma$~=~59~$\pm$~1~eV. This value was extracted from the fit of the spectrum using the sum of two Gaussian functions, one for the Mn $K_{\alpha}$ peak and one for the $K_{\beta}$ peak, imposing the same sigma for both peaks.}
    \label{fig:spectrumfenew}
\end{figure}
\begin{table}[ht!]
   \centering
    
    \begin{tabular}{l c c}

\hline 
    Mass & 3.5 & g \\ \hline
    Density & 5.32 & g/cm$^3$ \\ \hline
    Diameter & 5.08 & cm \\ \hline
    NTD response & 1645 & $\mu$V/MeV \\ \hline
    Baseline resolution (RMS)  & 44.5 $\pm$ 0.8 & eV \\ \hline
    Peak $\sigma$ at 5.9 keV  & 59 $\pm$ 1 & eV \\ \hline
 \end{tabular}
   \caption{Summary of the performance of GaAs-2 detector operated as a low-temperature calorimeter.}
   \label{tab:summarynew}
\end{table}

As reported in Tab.\ref{tab:summarynew}, the baseline resolution was measured to be 44.5~eV. Setting the threshold at 3$\sigma$ results in a threshold energy of 133.5~eV, corresponding to an improvement in resolution by approximately a factor of three with respect to our previous work~\cite{refId0}. Furthermore, Fig.~\ref{fig:spectrumfenew} clearly shows a distinct separation between the $K_{\alpha}$ and $K_{\beta}$ Mn peaks, highlighting the enhanced performance of the detector.

Since the $^{55}$Fe X-ray calibration source was deposited on an aluminum foil and in the low-energy region E < 5~keV, as shown in the inset in Fig.~\ref{fig:spectrumfenew}, we clearly identify the $K_{\alpha}$ peak of aluminum and a contribution from the $K_{\alpha}$ line of calcium.

The improved performance of the new GaAs crystal is likely related to its reduced mass, as in many bolometric detectors the performance scales with the absorber mass.

\section{Conclusions and perspectives}
The first successful measurements of the GaAs crystal as a cryogenic scintillating calorimeter with a dual readout (scintillation light and heat) were performed at the underground site of the Laboratori Nazionali del Gran Sasso (LNGS, Italy).

In the heat readout channel, the low-energy threshold of the 4.3~g GaAs detector was lowered from 1.5~keV to 360~eV, when compared to the previous measurements~\cite{refId0}. At the same  time, the energy resolution at 5.9~keV X-ray peak was measured as 140~$\pm~8$~eV. Both indicating significant improvement in noise and vibration reduction thanks to the latest modification of the IETI dilution refrigerator at LNGS.

In similar measurements with the 3.5~g GaAs detector equipped with the same Ge-NTD thermistor, the low-energy threshold was even further lowered to 133.5~eV, while 59~$\pm~1$~eV of energy resolution was achieved for a 5.9~keV X-ray peak.

The results obtained are highly encouraging in view of the development of low-energy threshold high-resolution cryogenic detectors based on GaAs crystals, aimed at direct low-mass dark matter searches. Especially, when focusing on the detection channel involved electron scattering or dark photon absorption.
Despite the current limitations imposed by the use of the Ge-NTD sensor for the heat channel readout, the observed detectors performance provides valuable insights for further optimising the GaAs detector design and experimental configuration.
 
Moreover, thanks to the simultaneously recording scintillation light with the cryogenic Ge-LD, the particle discrimination capability based on the different light response to the different type of irradiation has been demonstrated with the GaAs crystal for the first time. 

The light yield measured for $\beta$/$\gamma$ induced events was evaluated at 0.07~$\pm$~0.001~keV/MeV (0.05~photons/keV).

While the light yield estimated for $\alpha$ induced event exhibits 0.9~$\pm$~0.2~keV/MeV, i.e. a factor of 10 more than for $\beta$/$\gamma$ events. Such unusual luminescence properties of the GaAs crystal, i.e. producing more light when irradiated by $\alpha$ with respect to signals originated from $\beta$/$\gamma$s of the same energy, is similar to features observed with ZnSe and ZnO scintillating crystals, and is a subject for further detailed studies. Since it can lead to a highly sensitive detection method for low-energy nuclear recoils as a result of light-mass DM particles scattering on Ga or As nuclei.

To further improve the low-energy threshold in the heat channel of the GaAs detector there are few possible options. One option would be to apply the NTL effect to a GaAs detector.

Another option would be to adopt Transition Edge Sensors (TES) as thermal sensors, for both heat and light signal readout.

These advancements are expected to greatly enhance our measurement capabilities, extending the reach of direct dark matter detection and enabling the exploration of previously inaccessible regions of the dark matter parameter space.

------------
\section*{Acknowledgements} 
This work has been funded by the Italian University and Research Ministry thought the grant PRIN2022 - NextGenerationEU.
We are particularly grateful to Dr. J. Walker Beeman for his invaluable support in the  production of NTDs. We are also grateful to M. Guetti for the assistance in
the cryogenic operations and the mechanical workshop of LNGS. 

\section*{Availability of data and material}

Data will be made available on reasonable request to the corresponding author.

\bibliography{main.bib}
\bibliographystyle{spphys} 

\end{document}